\documentclass[reprint]{JASAnew}
\usepackage{epstopdf}
\usepackage{multirow}
\usepackage{setspace}

\newcommand*\patchAmsMathEnvironmentForLineno[1]{%
  \expandafter\let\csname old#1\expandafter\endcsname\csname #1\endcsname
  \expandafter\let\csname oldend#1\expandafter\endcsname\csname end#1\endcsname
  \renewenvironment{#1}%
     {\linenomath\csname old#1\endcsname}%
     {\csname oldend#1\endcsname\endlinenomath}}%
\newcommand*\patchBothAmsMathEnvironmentsForLineno[1]{%
  \patchAmsMathEnvironmentForLineno{#1}%
  \patchAmsMathEnvironmentForLineno{#1*}}%
\AtBeginDocument{%
\patchBothAmsMathEnvironmentsForLineno{equation}
\patchBothAmsMathEnvironmentsForLineno{align}%
\patchBothAmsMathEnvironmentsForLineno{flalign}%
\patchBothAmsMathEnvironmentsForLineno{alignat}%
\patchBothAmsMathEnvironmentsForLineno{gather}%
\patchBothAmsMathEnvironmentsForLineno{multline}%
}

\begin{document}
\title[Scattering statistics of rocks]{Scattering statistics of rock outcrops: Model-data comparisons and Bayesian inference using mixture distributions}
\thanks{Accepted to the Journal of the Acoustical Society of America. After it is published it will be found at \url{http://scitation.aip.org/JASA}}
\author{Derek R. Olson}
\email{dolson@nps.edu}
\affiliation{Oceanography Department, Naval Postgraduate School, Monterey, CA 93943}
\author{Anthony P. Lyons}
\affiliation{University of New Hampshire, Durham, NH}
\author{Douglas A. Abraham}
\affiliation{CausaSci, LLC, Ellicott City, MD}
\author{Torstein O. S{\ae}b{\o}}
\affiliation{Norwegian Defence Research Establishment, Kjeller N-2017, Norway}
\begin{abstract} 
The probability density function of the acoustic field amplitude scattered by the seafloor was measured in a rocky environment off the coast of Norway using a synthetic aperture sonar system, and is reported here in terms of the probability of false alarm. Interpretation of the measurements focused on finding appropriate class of statistical models (single versus two-component mixture models), and on appropriate models within these two classes. It was found that two-component mixture models performed better than single models. The two mixture models that performed the best (and had a basis in the physics of scattering) were a mixture between two K distributions, and a mixture between a Rayleigh and generalized Pareto distribution. Bayes' theorem was used to estimate the probability density function of the mixture model parameters. It was found that the K-K mixture exhibits significant correlation between its parameters. The mixture between the Rayleigh and generalized Pareto distributions also had significant parameter correlation, but also contained multiple modes. We conclude that the mixture between two K distributions is the most applicable to this dataset.
%

%
\end{abstract}
\maketitle

\section{Introduction}
The probability density function (pdf) of the amplitude of the acoustic field scattered by a rough, heterogeneous seafloor is closely tied to the properties seafloor environment. In acoustic target detection systems, it is intimately tied to system performance through the probability of false alarm, or pfa. For envelope-threshold detectors, the pdf is related to the pfa by $1 - F$, where $F$ is the cumulative distribution function (cdf) associated with the pdf. If the real and imaginary parts of the complex scattered pressure have a zero mean Gaussian pdf with equal standard deviations, then the pdf of the magnitude of the complex pressure (hereafter called the envelope pdf) will be Rayleigh, which is the case if the central limit theorem applies to the scattering process \cite{rayleigh_1919}.

Many measurements of the envelope statistics of seafloor scattering have resulted in envelope pdfs where where high amplitude compared to the mean events occur more frequently than the Rayleigh case, termed a heavy-tailed pdf \cite{crowther_1980, chotiros_etal_1985, gensane_1989,  stewart_etal_1994, lyons_abraham_1999, dorfman_dyer_1999,trevorrow_2004, becker_2004, lechenadec_2007, lyons_etal_2009,lyons_etal_2010,gavrilov_parnum_2010, gelb_etal_2010}. Several models have been used for heavy-tailed clutter statistics, including the Weibull (WB)\cite{schleher_1976}, log-normal (LN) \cite{chotiros_etal_1985}, K \cite{ward_1981}, and generalized Pareto (GP) \cite{lacour_2004} distributions. A few of these distributions have a basis in the physics of scattering, e.g. the K-distribution shape parameter is related to the effective number of scatterers within a resolution cell \cite{abraham_lyons_2002}.

Measurements of the envelope distribution from synthetic aperture sonar (SAS) images of glacially eroded rock outcrops off the coast of Norway are presented in this work in terms of the pfa\footnote{Although the pdf is a commonly analyzed quantity, the high amplitude low probability region (i.e. tail) is quite noisy using common estimation techniques. The pfa, which can be directly estimated from the data, is typically less noisy than the pdf}. These outcrops were selected because there is a paucity of scattering measurements from very rough surfaces, and glacially quarried surfaces have very large root mean square roughness \cite{olson_etal_2016}. For parts of the rock outcrop consisting of fractured surfaces, the image intensity changes rapidly over a variety of scales, and results in a heavy-tailed distribution. One of these outcrops was studied by \cite{gauss_etal_2015} in terms of moments of the scattered field and bathymetry; Heavy-tailed statistics were also observed.

These measurements were analyzed in two ways, 1) via model-data comparisons in terms of the Kolmogorov-Smirnov and Anderson-Darling tests, and 2) via analysis of the probability distribution of model parameters obtained using Bayes' theorem. It was observed that the pfa measurements are poorly fit by all of the models listed above. Mixture models, which consist of a linear combination of distributions, have also been used to model heavy tailed scattering \cite{gaullaudet_demoustier_2003,abraham_etal_2011,stanton_chu,lee_stanton_2014}, and we found that a good fit to the pfa is obtained using a two-component mixture. Our choice of a mixture model is motivated by the morphology of the outcrops studied in in this paper, which is presented in Sec.~\ref{sec:environment}. While the statistical tests give a good indication of model-data fit, the posterior probability distribution contains global information regarding the parameter space, including multiple modes and correlations between parameters.

Our long-term goal for this dataset is to link parameters of the environment to statistical parameters via a physical scattering model, and to perform geoacoustic inversion. Interpretation of the results will be performed with this orientation, although no geoacoustic inversion is performed here. We found that consideration of the connections between statistical parameters and environmental parameters is useful for determination of an appropriate statistical distribution. Relating model parameters to environmental parameters requires a physical scattering model that has been demonstrated to be accurate, eg. through numerical solution of the governing integral equations or scale model experiments. For the extremely rough surfaces studied in this work, such a model does not exist to date, and is an opportunity for future work.

In Sec. \ref{sec:environment}, an overview of the environment and acoustical measurements is given. The measured pfa from four outcrops is presented and compared to single component models in Sec.~\ref{sec:measurements}. In Sec. \ref{sec:modelAndInversion} the mixture model is presented, along with details of parameter estimation and inversion methods. We then present results and discuss them in Sec. \ref{sec:resultsAndDiscussion}, with conclusions given in Sec. \ref{sec:conclusions}.

\section{Environmental and acoustical data}
\label{sec:environment}
The roughness characteristics of the rock outcrops studied in this work were formed through glacial quarrying and abrasion. Quarrying results in large-scale roughness characterized by steps, and small-scale roughness on each step. Abrasion results in very low amplitude roughness at small scales, and large-scale smooth undulations due to the flow pattern of the glacier. Only acoustic measurements from quarried portions of the rock outcrops are examined in this work. More details on these classes of roughness, including roughness measurements of outcrops near the acoustic experiment using stereo photography can be found in \cite{olson_etal_2016} and references therein. These roughness characteristics are common in high-latitude environments where glacial erosion is the dominant force on the geomorphology.

Synthetic aperture sonar (SAS) images are the source of acoustic data in this work. SAS is a data collection and processing technique in which an acoustic transmitter and receiver array move along a track (which in this case was planned to be linear) and was traversed by an autonomous underwater vehicle (AUV). At regular intervals in space, pulses (linear frequency modulated in this case) are transmitted, and the scattered field collected along the receiver array. A diagram of the imaging geometry used in this study is shown in Fig. 6 of \cite{hansen_etal_2011}.

The measured data collected from multiple pings are combined into an array whose length is greater than the physical array, known as a synthetic array or aperture\cite{hawkins_thesis}. The synthetic aperture length is limited by the beamwidth of each receiver element, and thus the along-track resolution of the resulting image is proportional to the receiver element length. In practice, the sampling is nonuniform and the track is not straight due to oceanographic conditions. To correct for this, an inertial navigation system along with the displaced phase center (DPC) technique \cite{bellettini_pinto_2002} are used to estimate vehicle motion.

A matched filter is applied to each recorded ping in the frequency domain, and an additional apodization function is applied before transforming back to the time domain. The frequency apodization in this case was a Kaiser function \cite[p. 474]{oppenheim_schafer_dsp}, resulting in a peak side-lobe level of approximately -46 dB. After the matched filter, A synthetic aperture is formed for a given pixel by using a delay and sum beamformer for all the receive elements that have the pixel within the half power beamwidth. In practice only 75\% of this maximum aperture length is used to form the image, which reduces the effect of phase errors caused by residual unknown vehicle position at the expense of of resolution. Random phase errors along the synthetic aperture have been shown to produce increased side-lobe levels \cite{cook_brown_2009}. During the beamforming stage, a Hamming window \cite[p. 468]{oppenheim_schafer_dsp} is applied as a function of azimuth to control sidelobes in the along-track dimension, resulting in a peak sidelobe level of approximately -40 dB. The theoretical peak sidelobe levels in the along-track dimension functions are typically not achieved due to residual navigation errors.

\begin{figure*}
\figline{
\fig{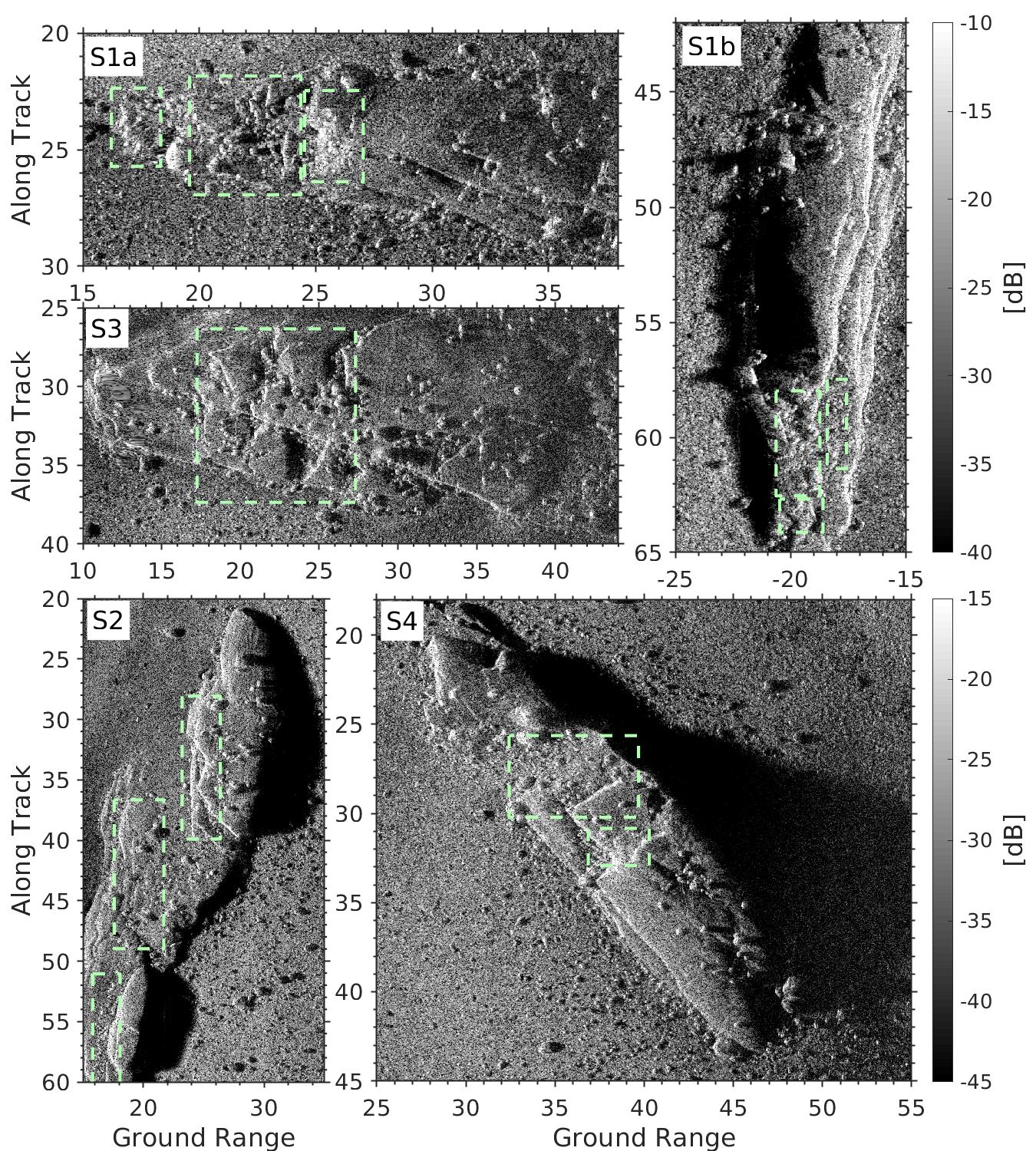}{.7\textwidth}{} \narrowcaption{.3\textwidth}{(color online) Synthetic aperture sonar images of rock outcrops. The horizontal axis denotes ground range (across track mapped to the seafloor using coarse SAS bathymetry estimates) position in meters. The vertical axis denotes the along track position in meters. Image intensity denotes the decibel version of the unaveraged dimensionless scattering cross section per unit area per unit solid angle. Boxes denote areas where pixels were extracted to form estimates of the pfa. These images are modified from \cite{olson_etal_2016}.\label{fig:sasImages}}}
\end{figure*}

The sonar used to collect data was the HISAS 1030, mounted on the HUGIN 1000 HUS AUV (both manufactured by Kongsberg Maritime) \cite{fossum_etal_2008}. The platform travelled at approximately 10 m altitude off of the sea floor. The system had a center frequency of 100 kHz, and a nominal bandwidth of 30 kHz. Typically, the achieved system resolution (when tapering and uncertainty in the vehicle navigation and soundspeed are taken into account) is approximately 3.3 cm in both the along track and across track directions. SAS images of four rock outcrops are shown in Fig.~\ref{fig:sasImages}, S1-S4. Images S1a and S1b are of the same rock outcrop from orthogonal imaging geometries. These images are modified from Fig. 10 in \cite{olson_etal_2016}, and were collected off the coast of Sandefjord, Norway in April 2011 \cite{midtgaard_etal_2011}. 

Image-level scattering statistics are studied in this work. Reverberation and image data are often spatially normalized (using, for example, a cell averaging constant false alarm rate (CA-CFAR) normalizer \cite{gandhi_kassam_1988}) prior to statistical analysis. This processing step is typically employed to remove the effects of smoothly varying changes, such as spatially varying terms in the sonar equation, or smooth changes in the environment. Most target detection systems use a CA-CFAR, or a similar type of normalizer. However, the focus here is on making inferences on the environment through the pdf, rather than system performance. Normalization may discard large-scale spatial information contained in the SAS image that may be important for environmental inference. Therefore, the pfa is estimated from calibrated image data but without any empirical normalization.

The SAS images used in this work are calibrated, meaning that intensity of each pixel corresponds to the unaveraged differential scattering cross section per unit area. The technique used to calibrate the system is detailed in \cite{olson_etal_2016}. It should be noted that the calibration was not based on laboratory measurements of the transmit level and receive sensitivity. Rather, it was based on comparisons between measured data and models produced using estimates of the geoacoustic properties and the roughness power spectrum of glacially abraded areas. Spatial changes in intensity due to the measurement system are absent, although pixel intensity is the result of averaging over the frequency and angular response of the imaging system. The sonar equation based calibration procedure used here (described in \cite{olson_etal_2016}) does not account for sources of nonstationarity due to the environment. Thus the intensity fluctuations in the image are due to changes in the local grazing angle, in seafloor composition, and in roughness properties.

\begin{figure}
	\includegraphics[width=0.5\textwidth]{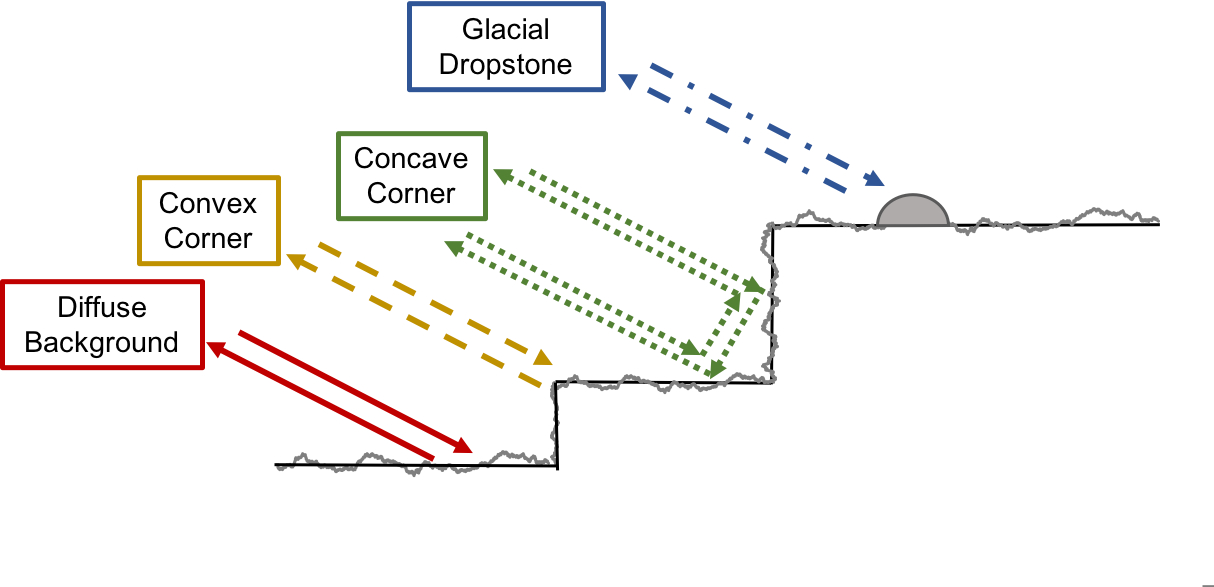}
	\caption{(color online) Diagram of scattering geometry. Each text box and set of arrows represent the geometry for the hypothesized scattering process responsible for the scattered field. Each pair of arrows denotes incoming and outgoing wave vectors for backscattering. The concave corner contains extra scattering paths due to multiple scattering. Dimensions in this figure are not meant to be realistic and are intended to be illustrative.}
	\label{fig:diagram}
\end{figure}

Glacially quarried regions studied in this work are denoted by dashed boxes in Fig.~\ref{fig:sasImages}. Examination of these regions shows that the scattered field potentially arises from four sources: low amplitude diffuse scattering from horizontally oriented facets, high amplitude near-specular scattering from vertically oriented facets (with multiple scattering from the concave corners likely), scattering from the convex edges, and scattering from glacial drop stones. An illustration of these scattering mechanisms in terms of the system and environmental geometry is shown in Fig.~\ref{fig:diagram}, in which dimension are not meant to be representative. The first of these components is lower amplitude, and distributed over large areas. The latter three components are higher amplitude and are distributed over smaller areas and are treated here as discrete scattering. Each arrow represents an incoming or outgoing ray path. The near-specular scattering paths are included in the diagram as part of the concave corner scattering paths.

To account for all of these sources of scattering, a four component mixture model would seem to be appropriate. If a Rayleigh distribution were used for each component, this model would have seven parameters, which can be considered to be the simplest four component model. Since clutter distributions commonly have two parameters, employing heavy-tailed distributions for the three discrete components would increase the number of parameters to eleven. 

However, the computational burden (for both time and memory) of a four component model with eleven parameters is prohibitive. The Bayesian inference performed in this work would have to be performed on a cluster, and statistical sampling of the parameter space would be required \cite[Ch. 11]{gelman}. More important, such a complex model would likely have a large degree of uncertainty in its parameter estimates. For reference, the four component Rayleigh mixture has an average relative parameter standard deviation (understood in terms of the Cramer-Rao lower bound, defined in Eq. (\ref{eq:crlb})), of 0.15, whereas the two component Rayleigh mixture has an average relative standard deviation of 0.040. Increasing the number of parameters from three to seven increased the uncertainty by about a factor of four. The average relative standard deviation of parameters for all the mixtures between Rayleigh and a clutter distribution (with four parameters) is 0.12. Increasing the number of parameters to eleven would likely increase the uncertainty to an unacceptable level. In this work, the sources of discrete scattering are combined into a single component, resulting in a simpler two component mixture.


%
%
%

\section{PFA Measurements and single component models}
\label{sec:measurements}
In this work, the pfa is estimated from a group of pixels defined by the dashed boxes in Fig.~\ref{fig:sasImages}. The complex image data within the boxes is decimated by a factor of nine in each dimension to ensure that the pixels represent independent samples. Although independent samples are not required to estimate the CDF (a binned histogram could be used), independent samples are required for any estimator based on the single-observation likelihood function, and for the evaluation of the Cramer-Rao lower bound below. The decimation factor was chosen by gradually increasing the decimation factor until the pfa shapes and parameter estimates began to converge. The decimated samples are then divided by the the average intensity. The pfa is estimated by $1 - F_{E}$, where $F_{E}$ is the empirical distribution function\cite{papoulis}.

\begin{figure*}
\figline{
\fig{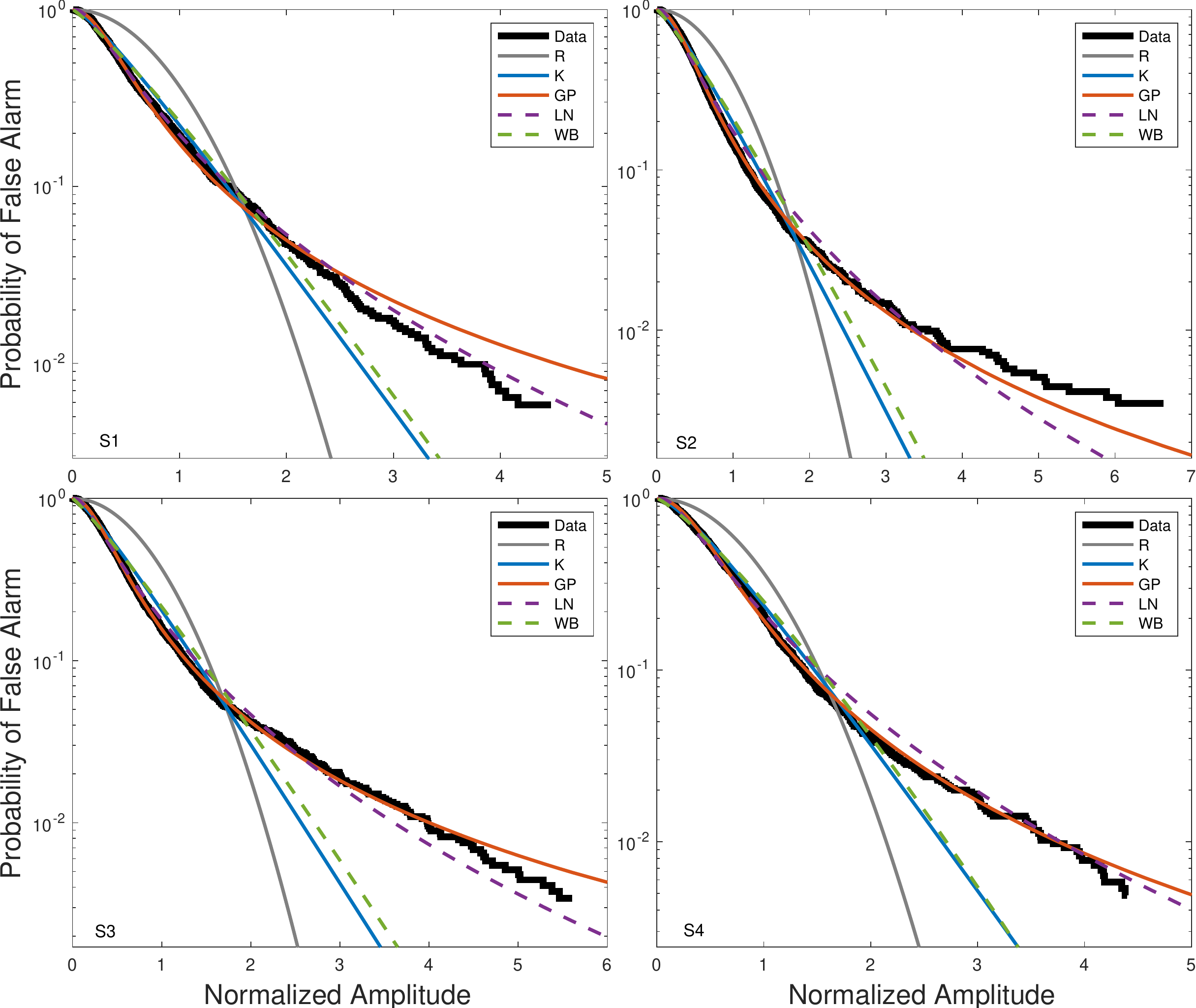}{.7\textwidth}{} \narrowcaption{.2\textwidth}{(color online) Plots of the pfa for each rock outcrop compared to single-parameter distributions on a log-linear scale.\label{fig:singleParameterModel}}}
\end{figure*}

Single component models were fit to the data, and plots of the model pfas are compared to the data in Fig.~\ref{fig:singleParameterModel}. In the pfa analysis outcrops S1a and S1b are combined into a single dataset by concatenating the decimated pixels before normalizing by the mean intensity. Each pfa is heavy-tailed and some have an inflection point, or knee, where there is a transition between concave and convex curvature in log-linear space (e.g. normalized amplitude of about 1.5 in Fig~\ref{fig:singleParameterModel}, in S1, and S2). Others change from concave downwards to flat, as in Fig~\ref{fig:singleParameterModel}, S3 and S4 near a normalized amplitude of 1.5 and 2 respectively. In this figure, pfa estimates below $10/N$, where $N$ is the number of samples, are not shown due to their high uncertainty, although they are used to compute parameter estimates and test statistics here and below. The number of samples obtained from each outcrop are 1722, 3143, 2926, 2054 for S1-S4 respectively.

\begin{table}
	\centering
	\begin{tabular}{l c  l l }
	\hline \hline
	Distribution &  pdf & Sh.  & Sc.  \\
	\hline
	Rayleigh & $2x/(\lambda) \exp\left(-x^2/\lambda\right)$ & -  & $\lambda$ \\
	K & $4/(\sqrt{\lambda}\Gamma\left(\alpha \right))( x / \sqrt{\lambda}  )^\alpha K_{\alpha-1}\left[\frac{2x}{\sqrt{\lambda}} \right]$ & $\alpha$ & $\lambda$ \\
	Weibull & $2 \beta x^{2\beta-1} / \left(\lambda^\beta\right) e^{-(x^2/\lambda)^\beta}$ & $\beta$ & $\lambda$ \\
	log-normal & $2/(x\sigma \sqrt{2\pi}) e^{-\left( 2\log(x/\lambda)\right)^2/(2\sigma^2)}$ & $\sigma$ & $\lambda$\\
Generalized       &\multirow{2}{*}{$2 x/\left({\lambda\left(1 + \gamma x^2 \lambda^{-1} \right)^{\gamma^{-1}+1}}\right)$}                      & \multirow{2}{*}{$\gamma$} &\multirow{2}{*}{$\lambda$} \\
Pareto &  & & 
	\end{tabular}
	\caption{Probability distributions used in this work. Mixture models are formed from linear combinations of these distributions. Columns starting from the left are the distribution name, probability density function, shape parameter, and scale parameter.}
	\label{tab:distributions}
\end{table}

The pdfs of the statistical models are given in Table \ref{tab:distributions}, and include the Rayleigh (R), K, Weibull (WB), log-normal (LN), and generalized Pareto (GP) distributions. In this table, $x$ is the pixel amplitude, $K_\nu(z)$ is the modified Bessel function of the second kind of order $\nu$, and $\Gamma(z)$ is the gamma function. The scale parameter for these distributions is $\lambda$, and the shape parameters are denoted $\alpha$, $\beta$, $\sigma$, and $\gamma$ for the K, WB, LN and GP pdfs respectively. Note that the GP pdf given is the equivalent envelope pdf when the GP distribution is used for the intensity. This choice was made because the GP distribution reduces to the exponential distribution when $\gamma=0$. The pdf for intensity, $y$, can be obtained using the substitution $x = \sqrt{y}$ and multiplying by the Jacobian $1/(2\sqrt{y})$.

Several of these distributions have a basis in the physics of scattering and can be derived from modulated Rayleigh-distributed clutter through a compound, or product model \cite{ward_1981},
\begin{align}
	p(x) = \int \textrm{d}\lambda\, p_{\lambda}(\lambda)p_R(x|\lambda)
	\label{eq:productModel}
\end{align}
where $p(x)$ is the pdf of the observed envelope, $p_R(x|\lambda)$ is the Rayleigh pdf with power $E[x^2] = \lambda$, and $p_{\lambda}(\lambda)$ is the distribution of the scale parameter, which is equal to the average intensity of the Rayleigh distribution. If $\lambda^2$ is gamma-distributed, then $p(x)$ will be K-distributed \cite{ward_1981,ward_tough_2002}. Similarly, if $\lambda^2$ has an inverse gamma distribution, then $p(x)$ will have a generalized Pareto distribution \cite{barnard_kahn_2008}. Other statistical models can also be generated from this model \cite{lyons_abraham_1999}, however, the log-normal and Weibull pdfs cannot.

Parameter estimates and test statistics for the single component models are shown in Table~\ref{tab:statParamsSingle}. For all cases, the maximum likelihood (ML) technique was used. Both the Anderson Darling upper tail (AU) test \cite{sinclair_etal_1990} and the Kolmogorov-Smirnov (KS) test \cite[p. 361]{papoulis} were used to test goodness of fit. The AU and KS test statistics are scaled between 0 and 1, in the same way as p-values. Since the parameters were estimated from the data they do not represent the true p-value. The Anderson-Darling test statistic emphasizes the upper tails of the distribution and is based on an average squared distance between the data and model CDF, whereas the KS test statistic weights all values of the CDF equally and is based on the maximum absolute CDF difference.
\begin{table}
\singlespacing
\centering
\begin{tabular}{c | l c c c c}
\hline\hline
  Set      & Dist. &    AU & KS  & Shape & Scale  \\
\hline
\multirow{5}{*}{S1} %
&R  & 0.000 & 0.000 & - & 1.000 \\ 
&K  & 0.000 & 0.000 & 0.790 & 1.008 \\ 
&LN & 0.000 & 0.000 & 1.835 & 0.207 \\ 
&WB  & 0.000 & 0.000 & 0.563 & 0.510 \\ 
&GP  & 0.223 & 0.405 & 0.989 & 0.214 \\ 
\hline
\multirow{5}{*}{S2}%
&R  & 0.000 & 0.000 & - & 1.000 \\ 
&K  & 0.000 & 0.000 & 0.854 & 0.802 \\ 
&LN & 0.000 & 0.000 & 1.762 & 0.191 \\ 
&WB  & 0.000 & 0.000 & 0.564 & 0.450 \\ 
&GP  & 0.311 & 0.247 & 0.811 & 0.222 \\ 
\hline
\multirow{5}{*}{S3}%
&R  & 0.000 & 0.000 & - & 1.002 \\ 
&K  & 0.000 & 0.000 & 0.776 & 0.937 \\ 
&LN & 0.000 & 0.000 & 1.826 & 0.186 \\ 
&WB  & 0.000 & 0.000 & 0.549 & 0.456 \\ 
&GP  & 0.583 & 0.514 & 0.944 & 0.198 \\ 
\hline
\multirow{5}{*}{S4}%
&R  & 0.000 & 0.000 & - & 1.000 \\ 
&K  & 0.000 & 0.000 & 0.939 & 0.893 \\ 
&LN & 0.000 & 0.000 & 1.734 & 0.252 \\ 
&WB  & 0.000 & 0.000 & 0.604 & 0.586 \\ 
&GP  & 0.078 & 0.025 & 0.791 & 0.299 \\ 
\hline \hline
\end{tabular}
\caption{Parameters and statistical test statistics of the single parameter models. From left to right, the columns are the rock outcrop, single parameter distribution, Anderson Darling upper tail scaled test statistic (AU), Kolmogorov Smirnov scaled test statistic (KS), shape parameter (if any), and scale parameter}
\label{tab:statParamsSingle}
\end{table}

For each outcrop, the only model whose scaled test statistic for the KS or AD test was greater than 0.001 was the GP distribution, which had KS statistics of 0.40, 0.25, 0.51, and 0.025 respectively, and AU statistics of 0.22, 0.31, 0.48, and 0.08 respectively. The better fit relative to the other single component models (but poor by absolute standards) that the GP model provides to S1-S3 is likely because the GP distribution is extremely flexible, and can fit data with inflection points in log-linear space. However, from visual inspection, it does not do a perfect job of fitting the pfa near the knee in S3, or the tails in S1. Note that the log-linear scale of the pfa plots shows model-data mismatch in a relative, rather than an absolute sense. Thus what appears to be high model data error at high amplitudes corresponds to rather small absolute error.

As discussed in \cite{abraham_etal_2011}, the single parameter models provide a good estimate of the pfa only when they intersect with the data. In other regions, the models provide a poor estimate of the pfa. This poor fit, along with the knee in the data motivate the use of mixture models.

\section{Mixture Model and Inversion Methods}
\label{sec:modelAndInversion}
\subsection{Mixture Models}
\label{sec:model}
The mixture models used in this work consist of two components, a background, $p_0$, and a clutter $p_1$ component. In the work of \citep{abraham_etal_2011}, a Rayleigh distribution was used exclusively for the background model, although in this work, a K distribution is also used. From each distribution, the mixture pdf is formed by \cite{titterington_etal_1985}
\begin{align}
		p(x \vert\boldsymbol{\theta} ) = \rho \, p_0( x \vert \boldsymbol{\theta}_0) + \left( 1 - \rho \right) p_1( x \vert \boldsymbol{\theta}_1),
		\label{eq:mixtureModel}
\end{align}
where $p(x \vert\boldsymbol{\theta} )$ is the pdf of the envelope, $x$, with parameter vector $\boldsymbol{\theta}$, $p_0( x \vert \boldsymbol{\theta}_0)$ is the background pdf, $p_1( x \vert \boldsymbol{\theta}_1)$ is the clutter pdf, and $\rho$ is the mixing proportion, which defines the ratio of samples corresponding to the background and clutter, and specifies the relative weight between the clutter and background distributions. The background parameter vector $\boldsymbol{\theta}_0$ consists of a single parameter, $\lambda_0$ for the Rayleigh distribution, and consists of two parameters, $\alpha_0$ and $\lambda_0$ for the K distribution. The clutter parameter vector consists of the two parameters for the clutter distribution, of which the first is the shape parameter, and the second is the scale parameter. The parameter vector, $\boldsymbol{\theta} = \lbrace\rho,\, \boldsymbol{\theta}_0,\,\boldsymbol{\theta}_1\rbrace$ consists of the mixing proportion, and the parameter vectors for both component distributions.

Only two component mixture models are considered here. Infinite component mixture distributions, such as the K-A distribution \cite{middleton_1999}, and the Poisson-Rayleigh Distribution \cite{mcdaniel_1993,fialkowski_etal_2004} have previously been used to model heavy-tailed scattering statistics, but are applicable to lower-resolution environments and systems where the scattering results from a coherent sum of background and spiky clutter \cite{ward_tough_2002}. In the present case, due to the high resolution of the sonar, and the low sidelobe levels, each acoustic sample can be identified with one or the other mixture component, and not their coherent sum.

In this work, the distributions from Table~\ref{tab:distributions} are used, specifically combinations of Rayleigh background with K, Weibull, log-normal and generalized Pareto clutter distributions (the same distributions used in \cite{abraham_etal_2011}), as well as a K background with a K clutter distribution, (used in \cite{ward_tough_2002,dong_haywood_2010,farshchian_posner_2010}). Using the $K$ distribution for the background component was motivated by the behavior of the S3 and S4 pfa curves, which tended to a straight line in log-linear space both above and just below the knee, and because of the variety of physical interpretations it has \cite{abraham_lyons_2002}. Given that the roughness of the horizontal facets appears uniform and homogeneous, it may seem strange that the low-amplitude background region does not behave like a Rayleigh distribution (which has quadratic rather than linear behavior in log-linear space). However, closer inspection of the horizontal components in Fig.~\ref{fig:sasImages} (eg. S3 at 20m ground range and 29m along track ), shows that the pixel intensity of the horizontal facets is not constant, but decreases slightly as ground range increases. This trend is likely due to the decrease of scattering strength with the grazing angle of the facet. Since the rock outcrops are several meters high, the local grazing angle is much smaller than it would be at the surrounding sea floor. Small changes in range therefore produce larger changes in the local grazing angle than would occur for a patch of seafloor surrounding the outcrop. Combining Rayleigh-distributed samples whose power changes deterministically can result in heavy-tailed statistics through Eq. (\ref{eq:productModel}), which has been used by \cite{lyons_etal_2010,lyons_etal_2016}.

Below, likelihood functions for the pdf as well as its components are defined. The likelihood function, $\ell(\boldsymbol{\theta}|x)=p(x | \boldsymbol{\theta})$ is equal to the pdf of the mixture model as a function of the parameters, with the data held constant. For statistically independent samples of $x$, the likelihood function is the product of the likelihood for each sample \cite{papoulis},
\begin{align}
	\ell(\boldsymbol{\theta}\vert\boldsymbol{X}) = \prod\limits_{n=1}^{N} \ell(\boldsymbol{\theta},X_n)
	\label{eq:IDlikelihood}
\end{align}
where $\boldsymbol{X}$ is the collection of samples of the random variable $x$ with elements $X_n$ with $n$ ranging from 1 to $N$. We will also use the likelihood functions of the mixture components, defined by 
\begin{align}
	\ell_0(\boldsymbol{\theta}_0\vert x) &= p_0(x \vert \boldsymbol{\theta}_0) & \ell_1(\boldsymbol{\theta}_1\vert x) &= p_1(x \vert \boldsymbol{\theta}_1).
\end{align}
The clutter to background power ratio (CBR) is an important parameter for characterizing a mixture distribution. The CBR is defined as the ratio between the mean intensity of the clutter component to the mean intensity of the background component, defined by.
\begin{align}
\mathrm{CBR} = \frac{\int\textrm{d}x\, x^2 p_1(x\vert\boldsymbol{\theta}_1)}{\int\textrm{d}x\, x^2 p_0(x\vert \boldsymbol{\theta}_0)}
\end{align}
 Based on the interpretation of $p_1$ as the clutter component, the CBR should be greater than 1 in intensity, and greater than 0 decibels (dB). Otherwise, the labels ``clutter'' and ``background'' should be exchanged. For all subsequent discussions and reported CBR parameters will use the dB version.

\subsection{Parameter estimation methods}
\label{sec:inversionMethods}
\subsubsection{Expectation Maximization}
Expectation maximization (EM) is a method that can be used to efficiently maximize the likelihood function for mixture models \cite{dempster_etal_1977}. If we are given a set of values of the envelope data $X_n$, then the parameters of the mixture model, Eq.~(\ref{eq:mixtureModel}), could be estimated by finding the parameter vectors that maximize the likelihood function. With a mixture model, a maximum likelihood (ML) estimator is frequently difficult to implement analytically because the mixture pdf is the sum of its component pdfs. The expectation maximization algorithm addresses this problem by maximizing an intermediate function that is the weighted log-likelihood for each component distribution. We define the intermediate function as $Q$, with the $Q$ functions for each component are defined as \cite[p440]{bishop}
\begin{align}
	Q_i(\boldsymbol{\theta}_i) &= \sum\limits_{n=1}^{N} W_{i,n} \log p_i( X_n | \boldsymbol{\theta}_i).
	\label{eq:qFunction}
\end{align}
and $Q = Q_0$ + $Q_1$. The values of $W_{i,n}$ are weights with values between zero and one that constitute a soft partition of the data. They specify the probability (or belief in Bayesian terminology) that sample $n$ belongs to distribution $i$, with $i=0$ for the background, and $i=1$ for the clutter component. In this work, $W_{0,n}$ are referred to as the background weights, and $W_{1,n}$ clutter weights. These weights are computed using
\begin{align}
	W_{0,n} &= \frac{\hat{\rho}\ell_0(\hat{\boldsymbol{\theta}}_0|X_n)}{\hat{\rho} \ell_0(\hat{\boldsymbol{\theta}}_0|X_n) + (1 - \hat{\rho})\ell_1(\hat{\boldsymbol{\theta}}_1|X_n)} \label{eq:weights0}\\
	W_{1,n} &= 1 - W_{0,n}
	\label{eq:weights1}
\end{align}
where the hatted variables denote parameter estimates.

From the previous three equations, the weights can be estimated given parameter estimates, and the Q function can be maximized to obtain new parameter estimates. The expectation maximization algorithm proceeds in two steps \cite{dempster_etal_1977}. In step one (expectation), the weights are estimated from Eqs.~(\ref{eq:weights0}), and (\ref{eq:weights1}). In step two (maximization), improved parameter estimation is performed. The two steps are repeated until the parameter estimates converge to a specified criterion. These two steps are guaranteed to increase the likelihood function with every iteration \cite{titterington_etal_1985}. One method to initialize the EM algorithm is derived in \citep{abraham_etal_2011}, and used in this work. For more details on EM in general see \cite{titterington_etal_1985} or Ch 9 of \cite{bishop}. 

\subsubsection{Bayesian Inference}
Bayes' theorem allows one to estimate the probability distribution of the parameter vector of a model, given data and prior information regarding the model parmeters. In Bayesian analysis, this distribution is called the posterior probability distribution (ppd), and is given by \cite{gelman}
\begin{align}
	p(\boldsymbol{\theta}\vert x) = \frac{p(x \vert \boldsymbol{\theta}) p(\boldsymbol{\theta})}{p(x)}
	\label{eq:bayesTheorem}
\end{align}
where $p(\boldsymbol{\theta}\vert x)$ is the ppd (i.e. the pdf of the parameters given the data), $p(\boldsymbol{\theta})$ is the prior pdf of the parameters, $p(x \vert \boldsymbol{\theta})$ is the conditional probability of the data, given the model, and is interpreted as the likelihood function, $\ell(\boldsymbol{\theta}|x)$. The probability of the data (also called the evidence) is $p(x)$, and is given by
\begin{align}
p(x) = \int\textrm{d}\boldsymbol{\theta}\, p(x\vert \boldsymbol{\theta}) p(\boldsymbol{\theta}) 
\end{align}
and can be interpreted as a normalizing factor that ensures that the integral of the ppd over the parameter space is equal to one. The prior distribution incorporates any known information about the parameters. In this work, uninformative bounded uniform priors are used.

Although there are several choices for minimally informative priors, we choose bounded uniform distributions whose endpoints are chosen by trial and error such that the 1D marginal ppd at the edges of the domain decays to at least 1/100 times the maximum 1D marginal. This is a somewhat arbitrary choice that removes the very low probability tails of the ppd. Since bounded uniform priors are used, the maximum value of the ppd, called the maximum \textit{a posteriori} (MAP) estimate, corresponds to the maximum value of the likelihood function (ML estimate) so long as it falls within the domain of support of the prior pdfs. Thus when comparing the best fit parameters, the Bayesian MAP can be taken as a surrogate for ML when comparing to results from the EM algorithm.

To select the bounds of the prior distributions, a direct numerical search on the likelihood function was performed with randomly chosen initial values. The random initial values were drawn from a uniform distribution between 0 and 1 for $\rho$ and $\gamma$, and 0 and 15 for all other parameters. The restriction on $\rho$ ensures that the pdf is positive, and the restriction on $\gamma$ is due to the fact that the GP only has a finite mean intensity, $E[x^2]$, for $\gamma<1$, and has lighter than Rayleigh tails for $\gamma<0$. The randomized initial parameters ensured that any multiple local maxima in the likelihood function are included in the domain of support of the prior distributions. When maxima with $\mathrm{CBR} > 0$ dB were found, the prior distributions were specified by increasing their limits until they satisfied the criterion stated above.

For several mixture pdfs, this randomly initialized ML search resulted in a local or global maximum parameter estimate whose CBR was negative. From examining the pfa data in Fig.~\ref{fig:singleParameterModel}, the Rayleigh distribution is inappropriate for modeling the upper tails of the distribution. Parameter estimates with $\mathrm{CBR} < 0$ dB were discarded. For the K-K mixture, the pdf is invariant to exchange of the clutter and background parameters along with replacing $\rho$ by $1 - \rho$. This behavior results from the symmetry of the K-K mixture and in turn results in symmetry in the ppd. If the two peaks based on the symmetry are well separated, then only the peak with positive CBR is used, due to the high-amplitude nature of clutter.

Once the bounds of the prior distributions were set, a numerical estimate of the ppd was found by computing the likelihood function exhaustively over the support of the uniform prior distributions and computing the evidence numerically. The ppd is typically a high dimensional function and is difficult to visualize. In practice, it is commonly visualized in term of each of its one and two dimensional joint marginal distributions. These are computed by integrating the full ppd over all but one or two elements of the parameter vector, $\boldsymbol{\theta}$, and examining all combinations.
\section{Results and Discussion}
\label{sec:resultsAndDiscussion}
In this section parameter estimates and comparisons are presented between measurements of the pfa to mixture models. The results of the global Bayesian inversion are also presented in terms of the 1D and 2D joint marginal ppds.

\subsection{Mixture model parameter estimates and model-data comparisons}
\begin{table}
\centering
\singlespacing
\begin{adjustbox}{width=\columnwidth}
\begin{tabular}{c | l c c c c c c c c}
\hline\hline
   Data      & Dist. &    AU & KS   & $\rho$& Bk.  & $\lambda_0$ & Cl. & $\lambda$ & CBR \\
      Set      & Name &     &    & & Sh. & & Sh.&  & (dB) \\
\hline
\multirow{5}{*}{S1} %
&R-K   & 0.337          & 0.340 & 0.447 & - & 0.172 & 0.762 & 1.989 & 9.456 \\ 
&K-K  & \textbf{0.695}  & 0.780  & 0.685 & 1.897 & 0.153 & 0.808 & 2.980 & 9.197 \\ 
&R-GP  & \textbf{0.964} & \textbf{0.947} & 0.370 & - & 0.105 & 0.700 & 0.540 & 12.335 \\ 
&R-LN  & 0.633          & \textbf{0.808} & 0.399 & - & 0.097 & 1.434 & 0.533 & 11.887 \\ 
&R-WB & 0.237           & 0.328   & 0.493 & - & 0.191 & 0.574 & 1.038 & 9.398 \\ 
\hline
\multirow{6}{*}{S2}%
&R-K      & 0.026          & 0.008 & 0.644 & - & 0.272 & 0.548 & 3.283 & 8.211 \\ 
&K-K      & \textbf{0.645} & 0.556 & 0.919 & 1.515 & 0.273 & 0.558 & 12.346 & 12.207 \\ 
&R-GP MAP & 0.548          & \textbf{0.828} & 0.033 & - & 0.013 & 0.762 & 0.247 & 18.988 \\ 
&R-GP EM  & \textbf{0.638} & 0.616 & 0.209 & - & 0.480 & 0.984 & 0.181 & 13.731 \\ 
&R-LN     & 0.606          & \textbf{0.691} & 0.455 & - & 0.354 & 2.063 & 0.187 & 6.452 \\ 
&R-WB.    & 0.036          & 0.021 & 0.663 & - & 0.287 & 0.481 & 0.906 & 8.323 \\ 
\hline
\multirow{5}{*}{S3}%
&R-K & 0.023           & 0.026 & 0.571 & - & 0.220 & 0.594 & 2.935 & 8.985 \\ 
&K-K  & \textbf{0.881} & \textbf{0.969} & 0.874 & 1.447 & 0.254 & 0.710 & 7.380 & 11.529 \\ 
&R-GP & \textbf{0.978} & \textbf{0.996} & 0.031 & - & 0.014 & 0.896 & 0.220 & 21.775 \\ 
&R-LN & 0.832.         & 0.889 & 0.332 & - & 0.306 & 2.020 & 0.196 & 6.926 \\ 
&R-WB & 0.031          & 0.032 & 0.593 & - & 0.234 & 0.508 & 0.966 & 9.047 \\ 
\hline
\multirow{6}{*}{S4}%
&R-K      & 0.053           & 0.027 & 0.475 & - & 0.340 & 0.714 & 1.996 & 6.222 \\ 
&K-K      & \textbf{0.908}  & \textbf{0.802} & 0.937 & 1.399 & 0.402 & 1.819 & 4.107 & 11.230 \\ 
&R-GP MAP & 0.852           & \textbf{0.977} & 0.102 & - & 0.046 & 0.662 & 0.401 & 14.148 \\ 
&R-GP EM  & 0.668           & 0.514 & 0.275 & - & 0.839 & 1.000 & 0.198 & $\infty$ \\ 
&R-LN    & \textbf{0.997}  & 0.904 & 0.347 & - & 0.606 & 1.907 & 0.216 & 3.427 \\ 
&R-WB     & 0.033           & 0.008 & 0.564 & - & 0.370 & 0.543 & 0.967 & 6.572 \\ 
\hline \hline
\end{tabular}
\end{adjustbox}
\caption{Parameters of the mixture distribution for each rock outcrop. From left to right, the columns are the scaled Anderson-Darling upper tail (AU) statistic, the scaled Kolmogorov-Smirnov (KS) statistic, mixture weight parameters, shape parameter for the background distribution, scale parameter for the background distribution, shape parameter for the clutter distribution, scale parameter for the clutter distribution, and the CBR in decibels}
\label{tab:statParams}
\end{table}

\begin{table}
\centering
\singlespacing
\begin{tabular}{c | l c c c c c}
\hline\hline
   Data      & Dist.   & $\rho$& Bk.  & $\lambda_0$ & Cl. & $\lambda$ \\
      Set      & Name  &       & Sh. &              & Sh.&            \\
\hline
\multirow{5}{*}{S1} %
&R-K   & 0.078 &-&   0.074  &  0.085  &  0.123\\ 
&K-K  & 0.123  &  0.266  &  0.363  &  0.2936  &  0.201\\ 
&R-GP  & 0.074  &-&  0.097  &  0.024  &  0.019 \\ 
&R-LN  & 0.227  &-&  0.125  &  0.063  &  0.154 \\ 
&R-WB &  0.065  &-&  0.063  &  0.035  &  0.099\\ 
\hline
\multirow{5}{*}{S2}%
&R-K      & 0.036 &-&   0.035   & 0.051  &  0.122 \\ 
&K-K      & 0.022 &   0.083  &  0.113  &  0.286 &   0.266\\ 
&R-GP & 0.576  &-&  0.791  	&  0.013  &  0.013 \\ 
&R-LN     &0.058  &-&  0.048  &  0.017  &  0.028\\ 
&R-WB.    & 0.031  &-&  0.034  &  0.031   & 0.091\\ 
\hline
\multirow{5}{*}{S3}%
&R-K & 0.042  &-& 0.042  &  0.058  &  0.101 \\ 
&K-K  & 0.055  &  0.114   &  0.162  &  0.554  &   0.308\\ 
&R-GP & 1.828  &-&  1.219  &  0.011  &  0.046\\ 
&R-LN & 0.087  &-&  0.071  &  0.016  &  0.025\\ 
&R-WB & 0.037  &-&  0.0310  &  0.028  &  0.085 \\ 
\hline
\multirow{5}{*}{S4}%
&R-K      & 0.089  & &  0.066  &  0.075   &  0.138 \\ 
&K-K      & 0.051  & 0.095  &  0.185  &  1.991  &  1.349 \\ 
&R-GP & 0.219  &-&  0.324 &   0.018  &  0.011\\ 
&R-LN    & 0.111  &-&  0.073  &  0.019 &   0.030 \\ 
&R-WB    & 0.058 &-&   0.056  &  0.036  &  0.091 \\ 
\hline \hline
\end{tabular}
\caption{Cramer-Rao lower bound for the mixture distribution parameters evaluated at the parameter estimates. From left to right the parameters are the mixing proportion, the background shape parameter, the background scale parameter, the clutter shape parameter, and the clutter scale parameter. The lower bound is presented as the square root of the lower bound of the variance divided by the parameter estimate.}
\label{tab:crlb}
\end{table}
\subsubsection{Results}
Parameter estimates are shown in Table~\ref{tab:statParams}, along with the CBR and the scaled output of the AU and KS tests. The top two KS and AU tests for each dataset are bolded. The entries are all MAP estimates unless the EM and MAP estimates differed significantly. If the two methods resulted in a difference in parameters greater than $0.1\%$, then both MAP and EM parameters are listed on different lines. Since EM is guaranteed to converge to a local maximum in the likelihood function, the EM and MAP parameter estimates should only differ if EM was initialized near a secondary mode. Indeed, when the EM algorithm was run with randomized starting parameters, it converged to the MAP estimates. Thus the comparison between EM and the MAP estimates is really a test of the initialization method.

\begin{figure*}
\figline{
\fig{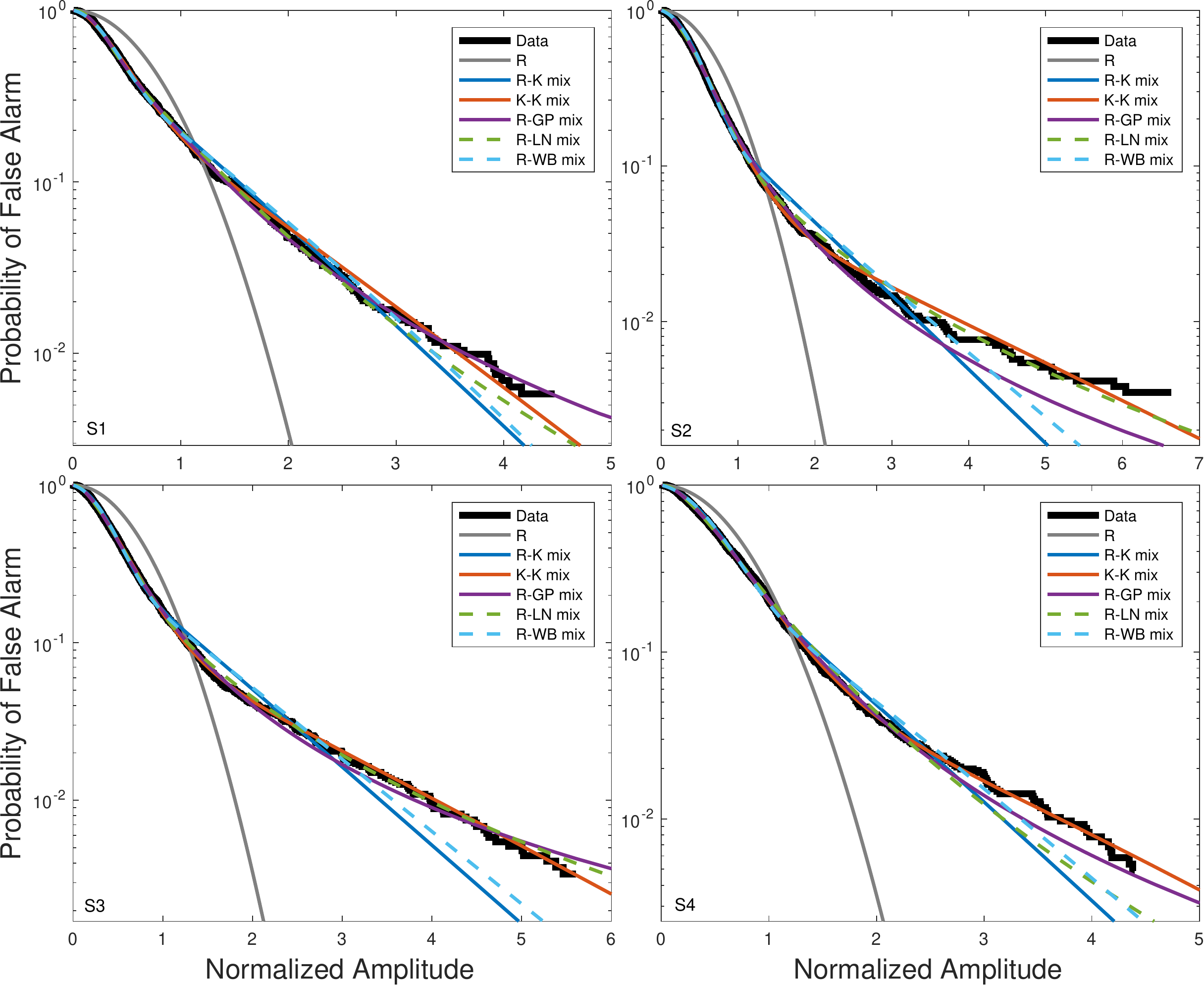}{.7\textwidth}{} \narrowcaption{.2\textwidth}{(color online) Plots of the pfa for each outcrop, along with mixture model results using MAP parameter estimates.\label{fig:mixtureModel}}}
\end{figure*}
Overall, the R-GP, R-LN and K-K mixture distributions perform the best in terms of the test statistics. Although the MAP estimate for the R-GP mixture has the best KS statistic for S2, the EM estimate has a higher AU statistic. This may be because the EM estimate has a higher shape parameter and is better able to fit the tails of the data. 

Plots of the pfa for rock outcrops along with the mixture models specified in \ref{sec:model} are shown in the subplots of Fig. \ref{fig:mixtureModel}. For all cases, the MAP estimates are used to compute model pfa curves. For all datasets, the R-GP, R-LN, and K-K mixture pdfs perform the best in terms a visual fit to the data, a confirmation of the analysis using the KS and AU statistics above. The R-K, and R-WB distributions perform poorly for all outcrops. For the models that perform well, the match is good at both high and low amplitudes. Many of the poorly performing mixture distributions have trouble fitting the tails of the distribution as well as the transition region between the background and clutter region (e.g. around a normalized amplitude of 2 in S3). 

\subsubsection{Discussion}
For the mixture pdfs used in this work, several of the parameters only have meaning in the context of that particular pdf. However, several parameters of mixture distributions have a strong relation to the environment, such as $\rho$ and CBR, although they may not strictly equal to environmental parameters. From Table~\ref{tab:statParams} these two parameters differ a great deal between each mixture model, and between each outcrop. This variability means that some of the models are likely more appropriate than others, although at the moment comparisons to ground truth cannot be made due to a lack of a valid scattering model. In the following, we propose several hypotheses to explain the variability in $\rho$. CBR estimates also varied a great deal, but we do not explore this quantity because we believe that it is related to the same sources of variability in $\rho$.

For the K-K mixture, $\rho$ tended to be high, and for the R-GP mixture it tended to be low. For the R-GP mixture and datasets S2 and S3, $\rho$ is extremely small, around 0.03. For other mixture distributions, $\rho$ tended to be around 0.4 - 0.6. Differences in estimates of $\rho$ between outcrops are plausible because the mixing proportion is related to the sizes of the geophysical features that cause background and clutter samples. From inspection of the SAS images, the background component appears to occupy a large number of samples in the image compared to the clutter component, at least above 0.5 for all images, but certainly larger than 0.7 or 0.8 for S2, S3 and S4. This very approximate estimate makes the $K-K$ mixture an attractive model because it agrees the most with qualitative inspection of the acoustic images.

To explain the variation, we could compare with ground truth in the form of SAS bathymetry estimates \cite{fossum_etal_2008,saebo_thesis}, which were available for these scenes. Quantitative comparisons are difficult to obtain from SAS bathymetry because of the presence of outliers in the height estimate, particularly for areas with low SNR, and the resolution is not high enough to resolve dropstones. For convex corners, the presence of multiple scattering makes estimating the phase between the two interferometric arrays difficult (due to the arrival times of each scattering path), and produces unreliable bathymetry estimates. Additionally, size of the specular returns in the image is not only dependent on the bathymetry, but is also dependent on the local grazing angle of the sonar, as well as the orientation of the nominally vertical facet. Multiple scattering from convex corners also causes a time delay in the scattered signals, causing multiple high amplitude bands to appear, or a smearing of the high amplitude region in the cross range direction if the resolution of the system is not high enough to resolve the multiple arrivals. Independent (i.e. non-acoustic) bathymetry estimates are required to perform this analysis. Therefore, a physics-based scattering model is absolutely required to be able to quantitatively compare SAS bathymetry with statistical modeling of SAS images.

One possible origin of the wide variation in estimates of $\rho$ is that this parameter is difficult to estimate precisely. This hypothesis can be tested using the Cramer-Rao lower bound (CRLB) for the various distributions. The CRLB gives a lower bound to the variance in the estimate of a parameter attainable when using an unbiased estimator (such as ML), and is related to the Fisher information matrix. For a given probability density function, $f(x|\boldsymbol{\theta})$, the elements of the Fisher information matrix, $\boldsymbol{I}\left( \boldsymbol{\theta}\right)$, are given by \cite{papoulis}
\begin{align}
\boldsymbol{I}\left( \boldsymbol{\theta}\right)_{i,j} = E\left[\left(\frac{\partial \log(f(x|\boldsymbol{\theta}))}{\partial \theta_i} \right) \left(\frac{\partial \log(f(x|\boldsymbol{\theta}))}{\partial \theta_j} \right)\right]
\end{align}
The CRLB for a single parameter, $\theta_i$, using an estimate computed with $n$ independent samples is given by 
\begin{align}
  \textrm{var}\left[\theta_i\right] \geq n^{-1}\left(\boldsymbol{I}^{-1}\right)_{ii}.
  \label{eq:crlb}
\end{align}
The Fisher information matrix was computed using numerical integration, due to the complexity of the mixture distributions. The CRLB is evaluated at the parameters estimated from the data, and gives an approximate indication of how easy or difficult is is to estimate each parameter.

Results for the CRLB are given in Table~\ref{tab:crlb}, and are presented in terms of the square root of the lower bound on the variance, i.e. the lower bound of the standard deviation, divided by the parameter estimate, in order to report the relative difficulty in estimating parameters. From the table, the relative CRLB for $\rho$ is generally quite small, with some exceptions in the K-K and R-LN mixtures for S1, the R-GP mixture for S2, the R-GP mixture for S4, and the R-GP and R-LN mixtures for S4. These mixtures have relative standard deviations of 0.1 or greater. Although some values of the CRLB are significant compared to the parameter estimate, we believe that the variance is not large enough to account for the significant variability in estimates of $\rho$ across distributions.

\begin{figure}
	\includegraphics[width=\columnwidth]{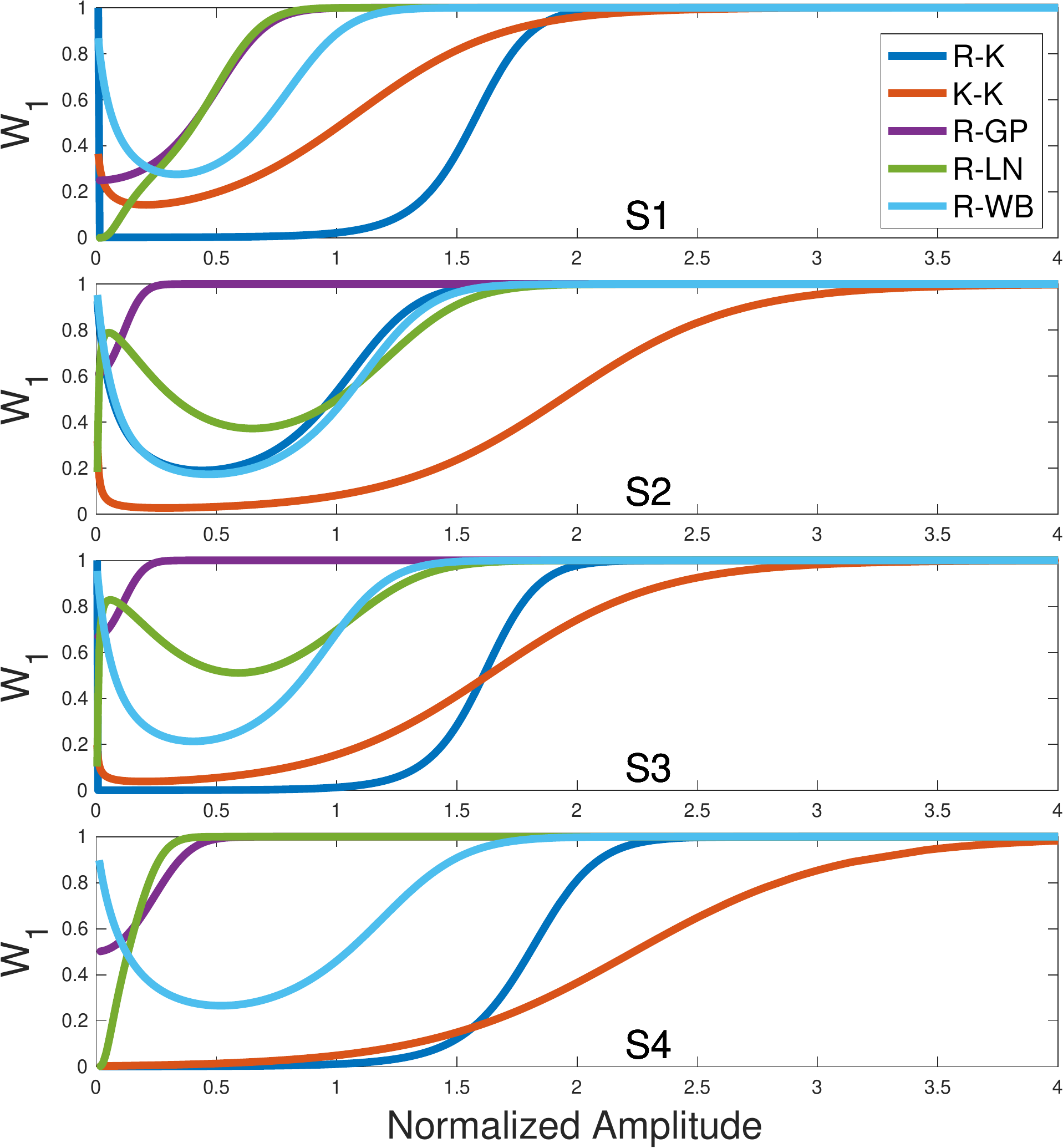}
	\caption{(color online) Belief weights from the EM algorithm for each mixture model and outcrop studied in this work.}
	\label{fig:beliefWeights}
\end{figure}

Another hypothesis is that the different clutter distributions can accurately model higher or lower amplitude samples from the ensemble, therefore assigning higher or lower values of the mixing proportion. The differing estimates of $\rho$ would then result from the properties of the distribution, rather than the environment. The EM algorithm can be used here as a diagnostic tool to test this hypothesis. The clutter weights, $W_{1,n}$, are used here to analyze whether samples are being assigned more weight to the background or clutter components. For each of the mixture distributions and data sets, $W_{1,n}$ is plotted in Fig.~\ref{fig:beliefWeights}. The weights were calculated using the ML parameter estimate, which is the same as the final EM estimate, so long as it achieves the global maximum. 

The R-GP distribution had systematically lower values of $\rho$ than all other distributions. These very small value for $\rho$ could result because the GP distribution can fit an inflection point in the log-linear amplitude pfa on its own, and with the R-GP mixture, the background component fits only the lowest amplitude data. The clutter belief weights for this distribution are quite high for much of the amplitude domain, and stay rather high even at low amplitudes. For this mixture, even very low amplitude pixels are assigned to have a clutter weight of about 0.3 for S1, and weights of 0.5 or above for all other outcrops. Based on Fig.~\ref{fig:beliefWeights}, we may conclude that the R-GP distribution tended to have low values of $\rho$ because most of the range of amplitudes can be accurately fit by the GP model. High values of the clutter weights for both low and moderate amplitude pixels does not make physical sense, and in our opinion counts as evidence against using the R-GP mixture to model these datasets.

For the $K-K$ mixture distribution, the opposite situation occurs, where the high amplitude pixels (above a normalized amplitude of say, 2.5), have a large weight assigned to the clutter distribution, and a small weight for the background distribution. At low amplitudes, the clutter weights are quite low, with only the R-K mixture (which did not fit the data very well) being lower. There is a region in which both the background and clutter components will contribute significantly to the total. The $K-K$ mixture has the lowest clutter belief weights at all amplitudes (apart from the R-K mixture), and therefore has the largest mixing proportion. The behavior of the K-K mixture weights makes physical sense: it assign low clutter weight to low amplitudes, high clutter weight to high amplitudes, and in the intermediate region, transitions between the two. This agreement with our qualitative inspection of the images gives a great deal of evidence in favor of the K-K mixture as the most appropriate mixture model out of the ones studied here. Based on the clutter weights in Fig.~\ref{fig:beliefWeights}, the variability in $\rho$ between each model is a direct result of the way that the mixture distributions partition the pixel amplitudes.

\subsection{Posterior Probability Distribution}
Due to the large amount of information present in the Bayesian inversions, only two datasets, S3 and S4 were analyzed, and only the K-K and R-GP mixture distributions were considered. This choice was made because the K-K and R-GP distributions both generally fit the measurements very well, and the distribution components have a basis in the physics of scattering. The log-normal distribution also fits these data very well but does not have a direct link to scattering physics.

\begin{figure*}
\figline{\fig{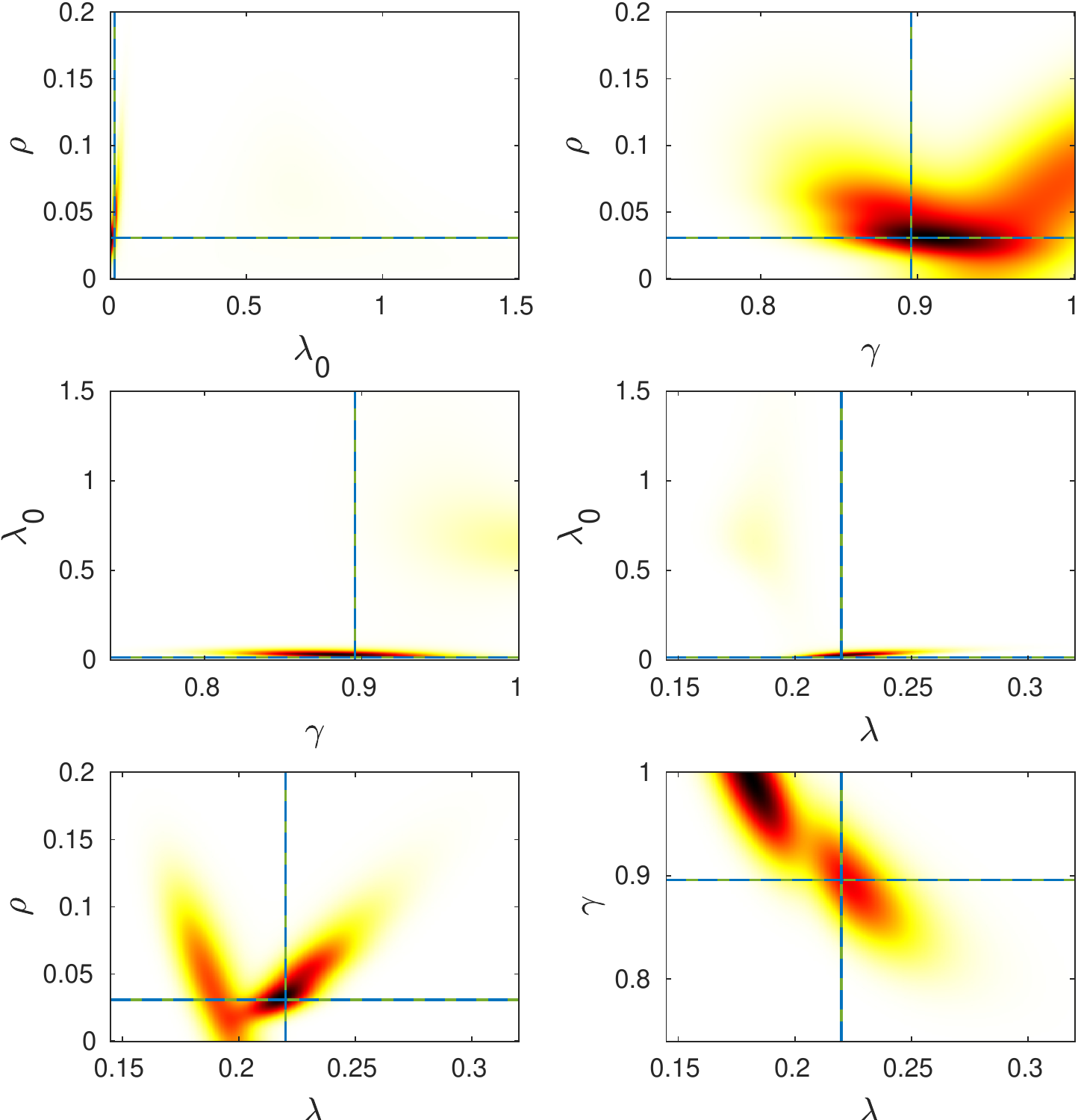}{.4\textwidth}{(A)}
\fig{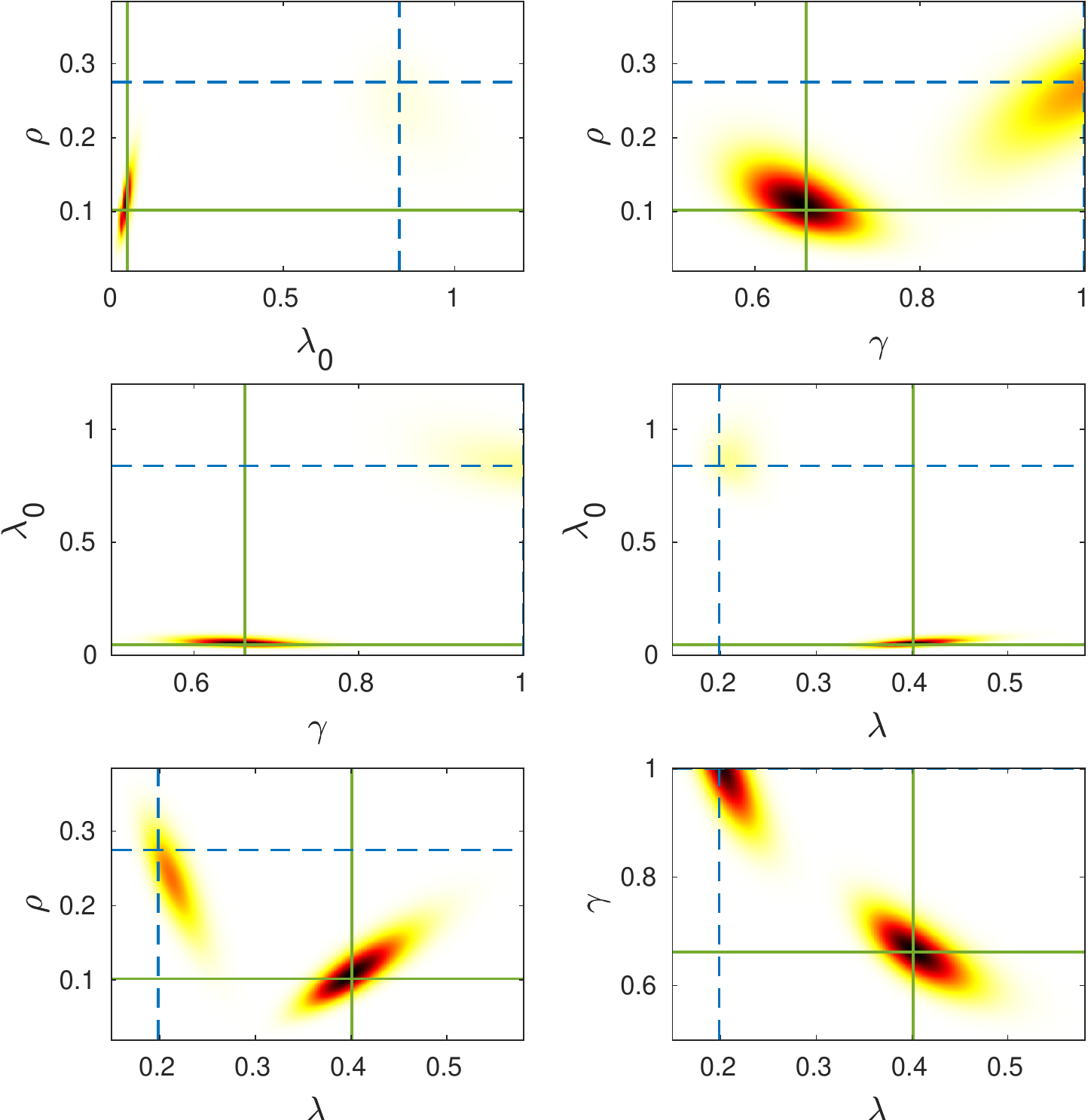}{.4\textwidth}{(C)}
}
\figline{\fig{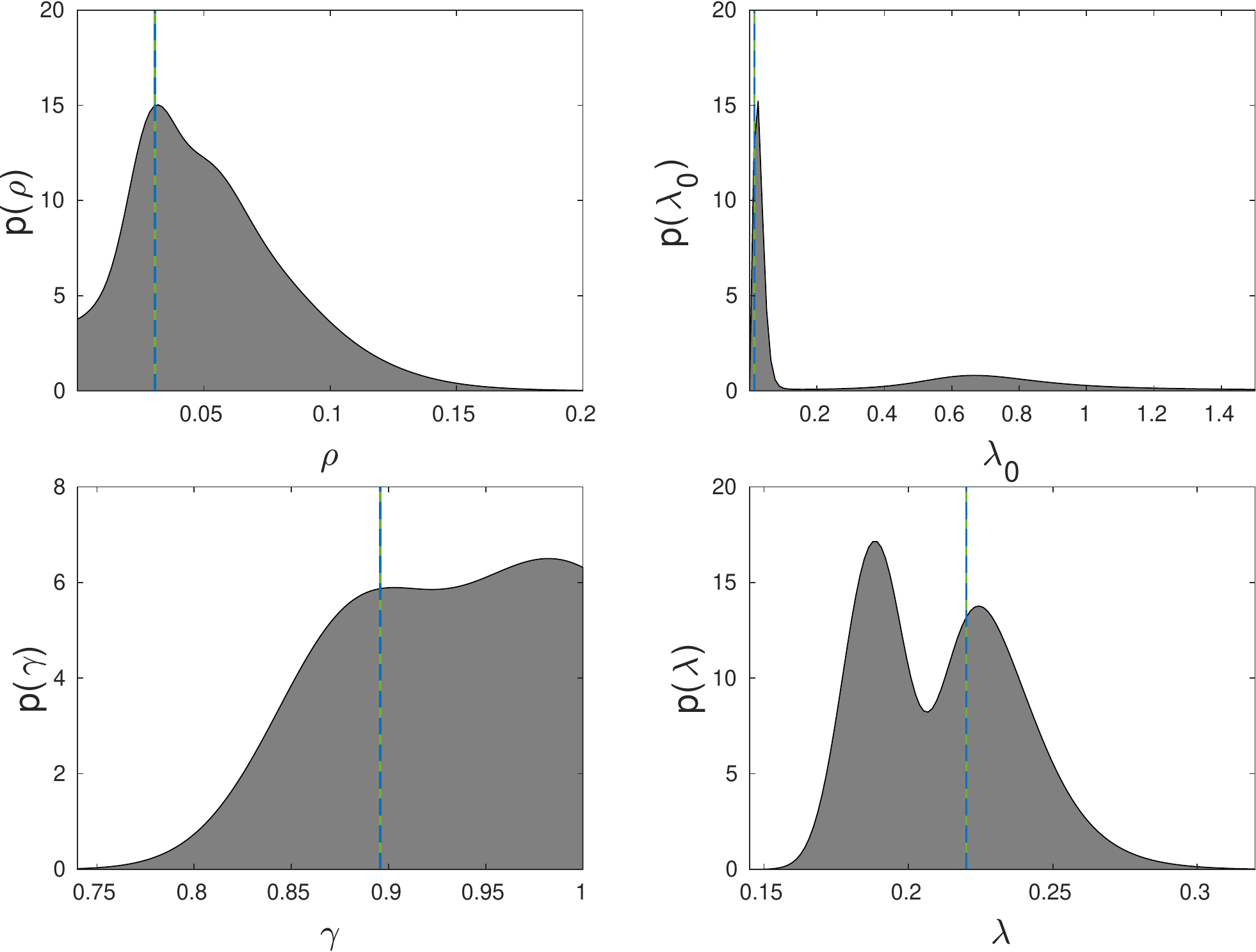}{.4\textwidth}{(B)}
\fig{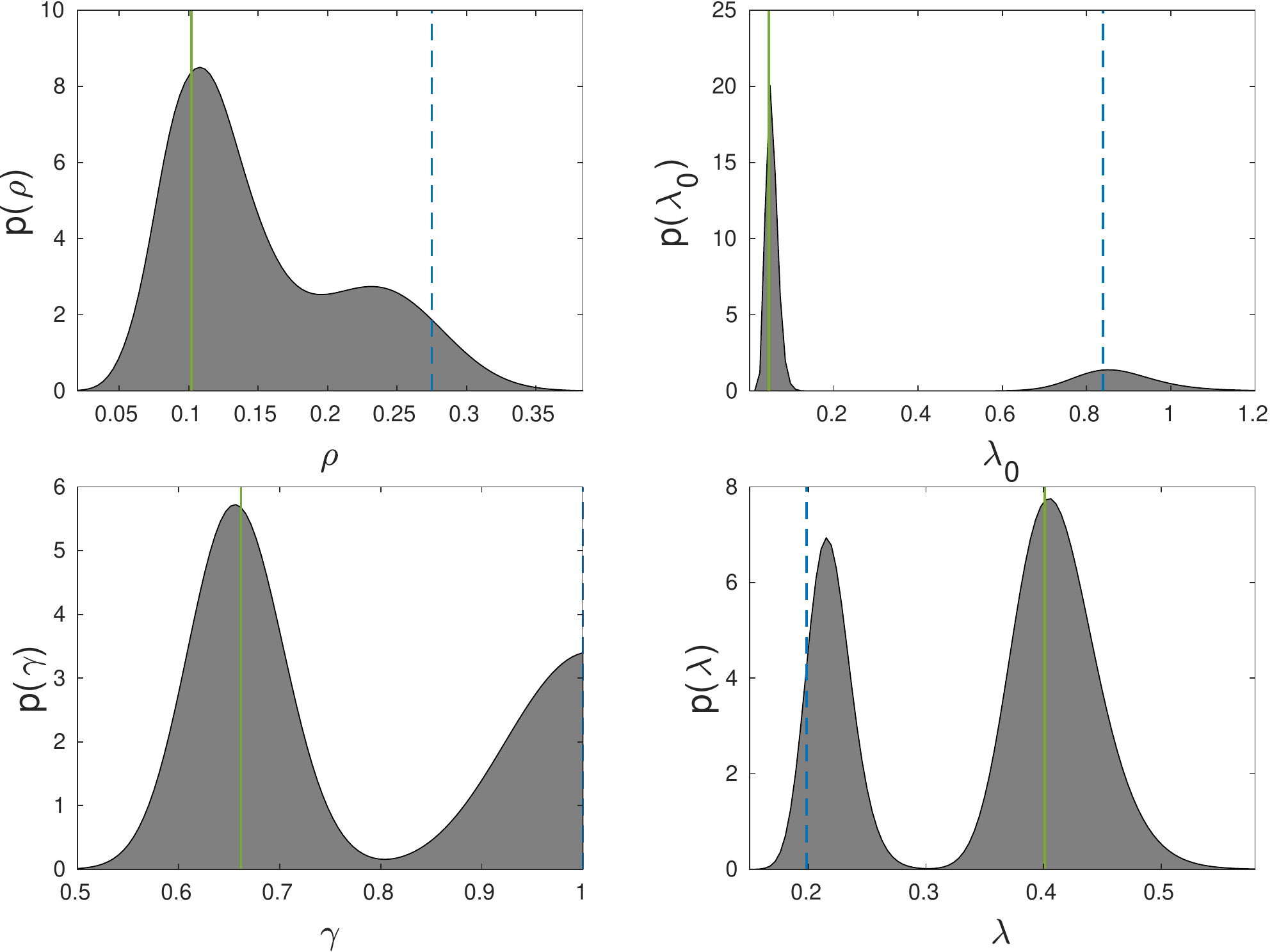}{.4\textwidth}{(D)}
}
\caption{(color online) Marginal distributions for the R-GP mixture distribution for outcrops 3 (A and B), and 4 (C and D). Solid lines indicate MAP parameter estimates, while dashed lines indicate EM. \label{fig:gpmarginals}}
\end{figure*}

The marginal distributions for the R-GP mixture are plotted in Fig.~\ref{fig:gpmarginals}. Solid lines denote the MAP parameter estimate, and dashed lines denote the EM estimate. The most striking feature of these marginal distributions is the presence of multiple local maxima, or multiple modes. Only the R-GP mixture distributions resulted in solutions with multiple local maxima in the ppd and did so for outcrops S2-4. This behavior is likely due to the extreme flexibility of the GP distribution. For inversion purposes, this extreme flexibility is a liability because it can increase parameter uncertainty, and obscures connections between statistical parameters and environmental parameters.

For all of the 2D marginal distributions in this figure, there is significant correlation between all of the parameters. Moreover, the shapes of the 2D marginal distributions indicate that the inverse problem is highly nonlinear. Inference for the physical properties of the rock outcrops using the R-GP mixture model would likely be difficult due to the multimodal behavior of the ppd and significant parameter correlation. Additionally, the GP distribution has infinite intensity variance for $\gamma>1/2$ \cite{hosking_wallis_1987}. Since the MAP estimate of the GP shape parameter falls within this regime for all datasets, the scintillation index (intensity variance divided by the intensity mean) of this model is also infinite. This is a nonphysical result (since the sample scintillation index is always finite), and gives further evidence against the suitability of the R-GP mixture for this dataset.
\begin{figure*}
\figline{\fig{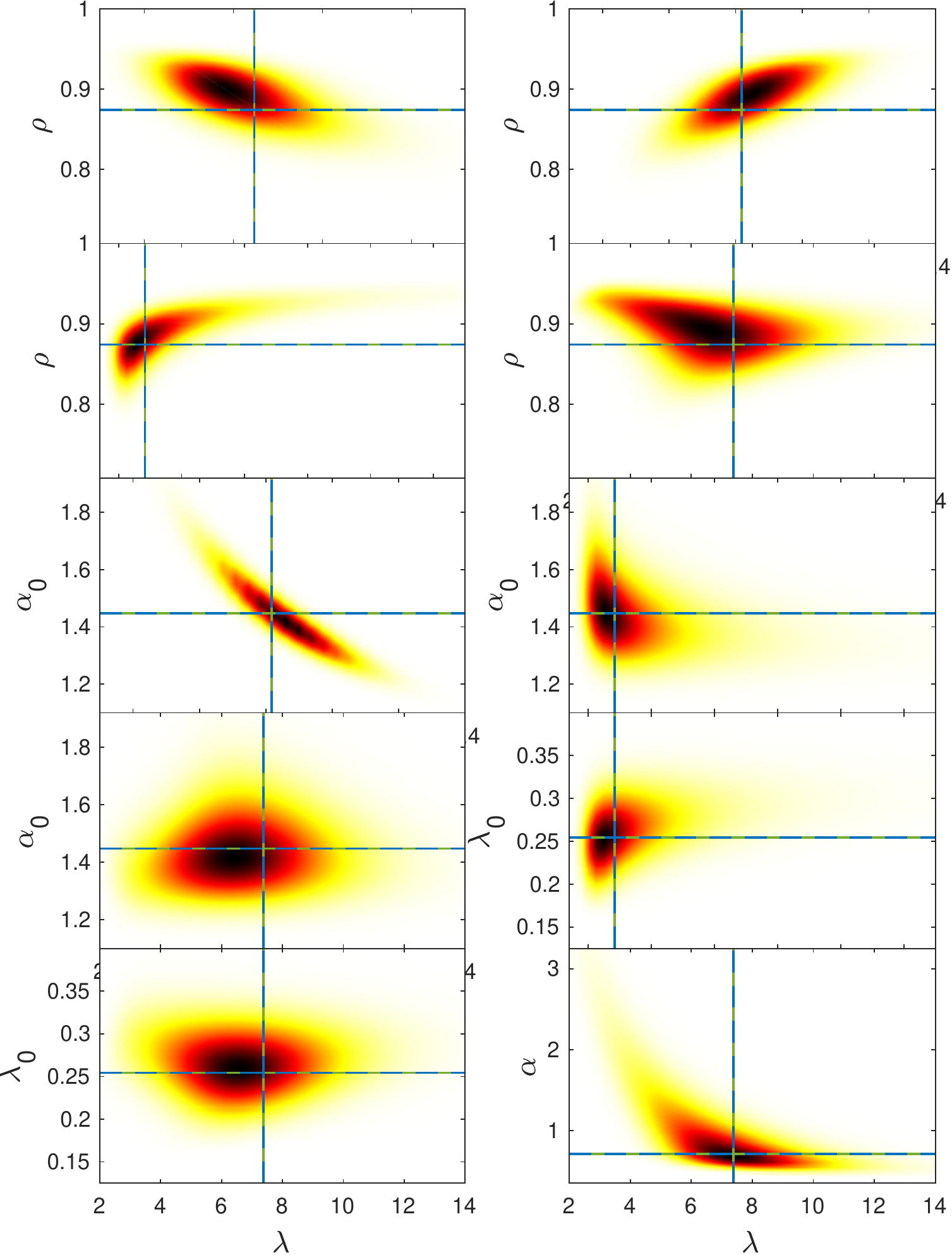}{.4\textwidth}{(A)}
\fig{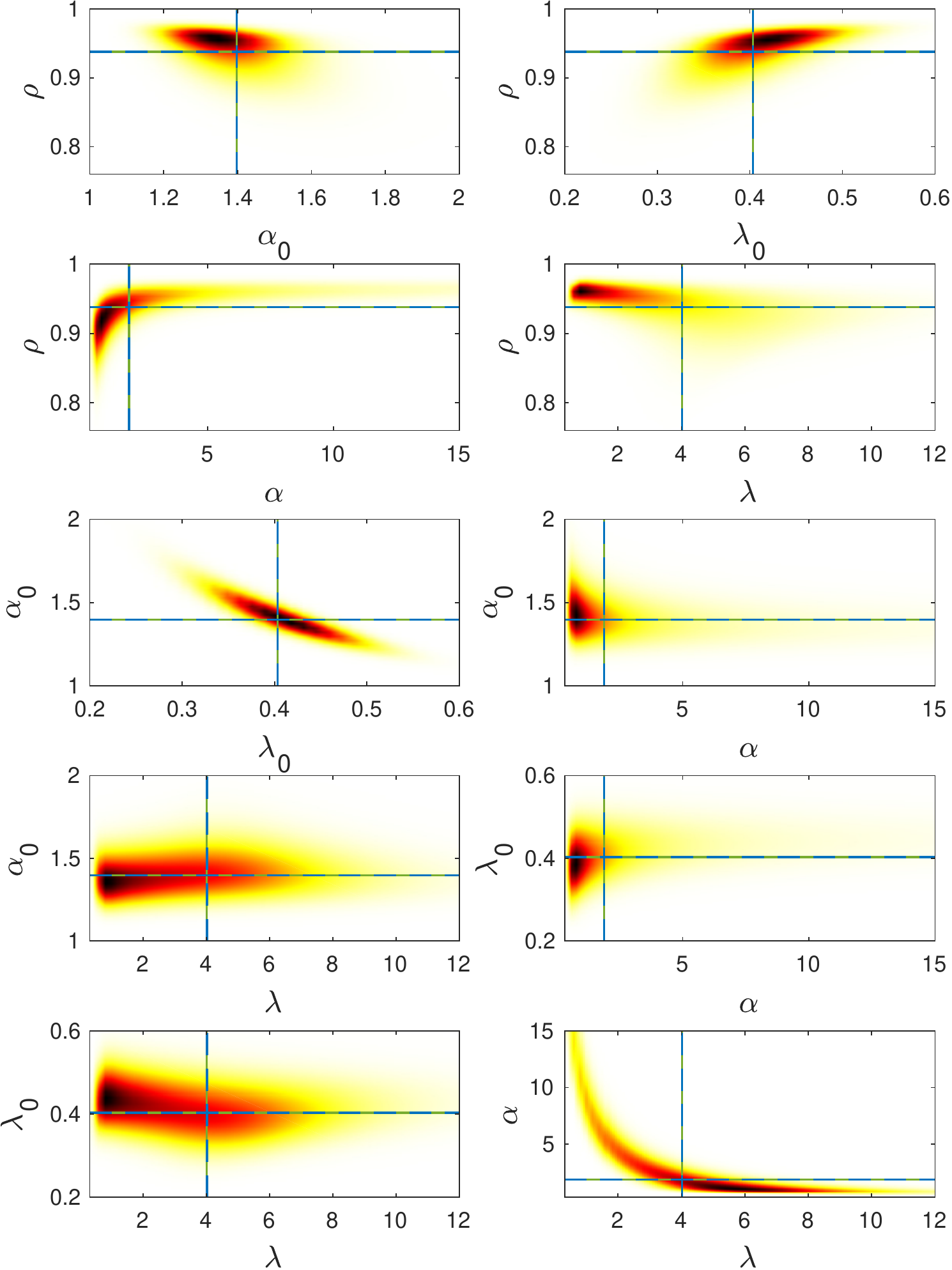}{.4\textwidth}{(C)}
}
\figline{\fig{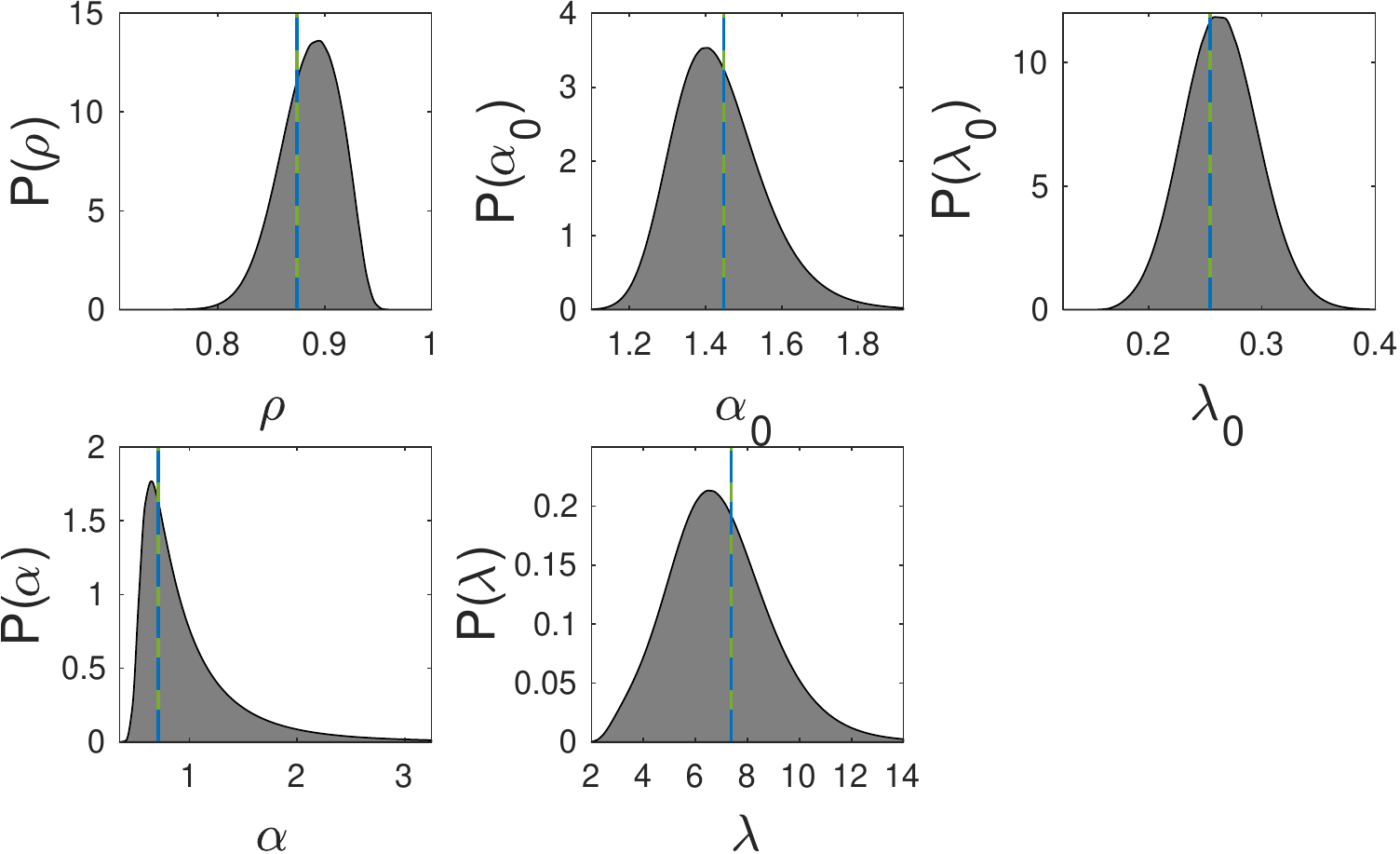}{.4\textwidth}{(B)}
\fig{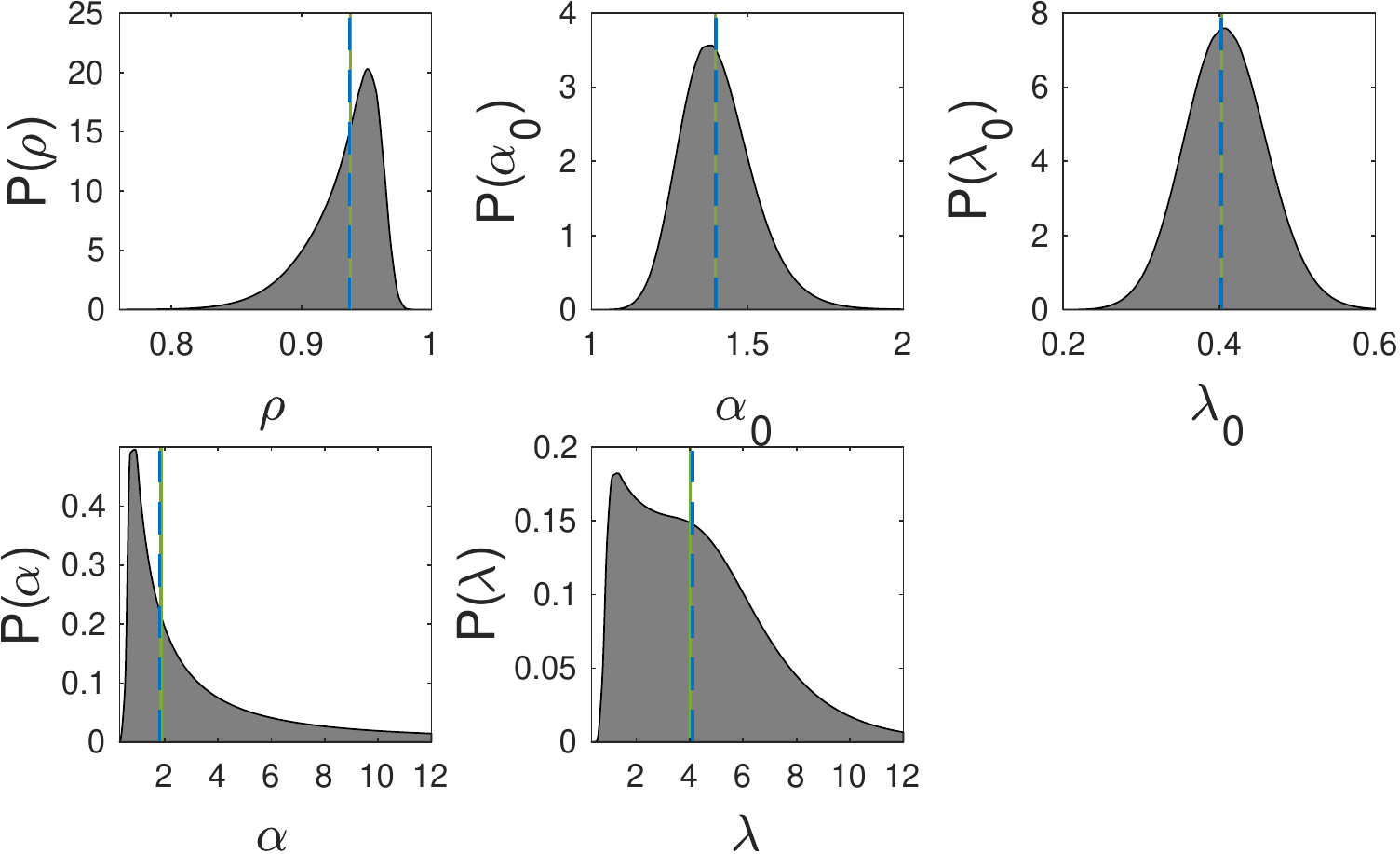}{.4\textwidth}{(D)}
}
\caption{(color online) Marginal distributions for the K-K mixture distribution for outcrops 3 (A and B), and 4 (C and D). Solid lines indicate MAP parameter estimates, while dashed lines indicate EM.\label{fig:kkmarginals}}
\end{figure*}

The marginal distributions for the K-K mixture are plotted in Fig.~\ref{fig:kkmarginals}. Unlike the R-GP mixture, the 1D and 2D marginals do not show multiple maxima, apart from the ambiguity between clutter and background due to the symmetry of the K-K mixture. The K-distribution is less flexible in terms of the shapes of the tails. As $\alpha$ tends to zero, the tails of the K pfa tend to a straight line in log-linear space, and as $\alpha$ tends to infinity, it approaches the Rayleigh case. Although the K distribution can have positive curvature in log-linear amplitude space for very small shape parameters, this only occurs at low amplitudes. This limitation on the  possible shapes that the K distribution can adequately fit provides some measure of protection agains multiple modes in the ppd.

Like the 2D marginals of the R-GP mixture, the K-K mixture exhibits strong parameter correlation. Even though parameter correlation is high, the high probability regions of several 2D marginals sometimes exhibit simple behavior, such as in $\rho$ versus $\alpha_0$ and $\rho$ versus $\lambda_0$ in both datasets. Other 2D marginals exhibit much more complex shapes such as curved peaks in $\alpha_0$ versus $\lambda_0$ and $\alpha$ versus $\lambda$ in both datasets. These two specific parameter correlations are likely the result of the K distribution having constant intensity along lines of constant $\alpha\lambda$ and $\alpha_0\lambda_0$. Like the R-GP distribution, inference regarding the physical properties of the rock outcrops using the scattered field statistics would likely be difficult due to parameter correlation, although the K-K mixture does not exhibit multimodal behavior.
\section{Conclusions} 
\label{sec:conclusions}
Measurements of the statistics of scattering from rock outcrops have been presented. Comparisons with several commonly used single-component statistical models fail to adequately model these data, due an inflection point in the PFA curve. Mixture models have an interpretation in terms of the physics of scattering from this environment and were found to provide a better fit to the data. It was found that the Rayleigh-Generalized Pareto, Rayleigh-log normal, and K-K mixtures performed the best in terms of the AU and KS test. Of those three, the mixing proportion of the K-K mixture corresponded the best with qualitative interpretation of the sonar images. This analysis was confirmed by analysis of the belief weights from the EM algorithm. A multimodal parameter ppd was observed in the R-GP mixture, as well as significant parameter correlation. For the K-K mixture, no multimodal behavior was observed, but significant parameter correlation was found. Both multimodal ppds and nonlinear parameter correlation pose significant problems if mixture models are to be used to perform geoacoustic inversion. Based on this analysis, the K-K mixture has the greatest evidence in its favor: 1) it performs well in terms of the AU and KS statistics, 2) its values of $\rho$ agreed the most with qualitative interpretation of the sonar images, 3) its clutter belief weights make intuitive sense based on the images, 4) it has a well-behaved ppd (i.e. a single maximum), and 5) the K distribution has links to the physics of scattering, both for discrete scatterers, \cite{abraham_lyons_2004}, and for homogeneous environments at high resolution \cite{lyons_etal_2016}.

\begin{acknowledgments}
\label{sec:acknowledgements}
\end{acknowledgments}
The authors would like thank to the crew of the HU Sverdrup II, operators of the HUGIN AUV, and researchers at the Norwegian Defence Research Establishment for conducting the acoustic scattering experiment and collecting the data. This work was supported by the U.S. Office of Naval Research under grants N00014-18-WX00776, N00014-16-1-2335, N00014-13-1-0056 and N00014-12-1-0546. The Hamming high performance computing cluster at the Naval Postgraduate School was used to perform some of the Bayesian inference calculations.

\end{document}